\documentclass[jcp,twocolumn]{revtex4-1}
\usepackage{amsmath}
\usepackage[english]{babel}
\usepackage{latexsym}
\usepackage{mathrsfs}
\usepackage{graphicx}
\usepackage{graphics}
\usepackage{color}
\usepackage{bm}
\usepackage{units,nicefrac}

\begin{document}

\title{Formation of Regression Model for Analysis of Complex Systems\\ Using Methodology of Genetic Algorithms}

\author{Anatolii~V.~Mokshin$^{1}$, Vladimir~V.~Mokshin$^{2,3}$ and Diana~A.~Mirziyarova$^{1}$}

\affiliation{$^1$Institute of Physics, Kazan Federal University,
      \it 420008 Kazan, Russia \\
   $^2$Institute for Computer Technologies and Information
   Protection,  Kazan National Research Technical University named after A.N. Tupolev-KAI,
   \it 420111 Kazan, Russia \\
   $^3$Almetyevsk branch of Kazan National Research Technical University named after A.N. Tupolev-KAI,
   \it 423461 Kazan, Russia \\
   {\small \rm E-mail:  anatolii.mokshin@mail.ru}}

\begin{abstract}
This study presents the approach to analyzing the evolution of an arbitrary complex system whose behavior is characterized by a set of different time-dependent factors. The key requirement for these factors is only that they must contain an information about the system; it does not matter at all what the nature (physical, biological, social, economic, etc.) of a complex system is. Within the framework of the presented theoretical approach, the problem of searching for non-linear regression models that express the relationship between these factors for a complex system under study is solved. It will be shown that this problem can be solved using the methodology of \emph{genetic (evolutionary)} algorithms. The resulting regression models make it possible to predict the most probable evolution of the considered system, as well as to determine the significance of some factors and, thereby, to formulate some recommendations to drive by this system.  It will be shown that the presented theoretical approach can be used to analyze the data (information) characterizing the educational process in the discipline ''Physics'' in the secondary school, and to develop the strategies for improving academic performance in this discipline.
\end{abstract}



\maketitle

\section{Description of the Approach \label{sec: method}}

\subsection{Factors and non-linear regression model.}

Let evolution of a complex system be characterized by a finite set of time-dependent factors. By \emph{external factors}, we mean those factors that are affected by the system. Then, the \textit{internal factors} are those factors that contain information generated by the system itself. Using external factors, it is possible to drive the considered system by setting the values of these factors. Internal factors characterize the response of the system. Therefore, their values can only be adjusted through external factors. To clarify this, we give the following example from physics.
Let some polymer material undergo external mechanical stress, namely, shear deformation.
In this case, the force causing the shear deformation can be considered as an external factor. The values of this force can be adjusted. In turn, as a result of the shear deformation, some structural changes in this system may occur~\cite{tempbib6}. These structural changes will be characterized by the internal factors:  local density, local temperature, and local tangential stresses. In the general case, this situation corresponds to a nonequilibrium physical process. Therefore, the relationship between these factors cannot be described in terms of any currently known physical laws; these relationships can be nontrivial~\cite{tempbib5,tempbib20}.

In accordance with the standard methodology of regression analysis~\cite{tempbib3}, it is necessary to introduce the concepts of \textit{input} and \textit{output} factors. By an \textit{output factor} we mean a factor by which the evolution of the system is monitored. It is necessary to note that the number of the output factors may be more than one, but it must be finite. All other factors $x_{1}(t)$, $x_{2}(t)$, $x_{3}(t)$, $\ldots$, $x_ {M}(t)$ will be defined as \textit{input factors}. Here, $M$ is the number of the input factors. The choice of an output factor is determined by the purpose of the study. Therefore, the attribution of the factors to input or output factors may be a matter of convention.

Let there be a relationship between the output factor $y(t)$ and the input factors
$x_{1}(t)$, $x_{2}(t)$, $x_{3}(t)$, $\ldots$, $x_{M}(t)$, which is defined as the generalized Kolmogorov-Gabor polynomial~\cite{tempbib7}:
\begin{eqnarray} \label{eq_general_polinom}
    y(t)= a_{0}+\sum_{i=1}^{M} a_{i} x_{i}(t)+ \sum_{i=1}^{M}\sum_{l>i}^{M}a_{il} x_{i}(t) x_{l}(t)+ \nonumber  \\
    \sum_{i=1}^{M}\sum_{l>i}^{M}\sum_{k>l}^{M}a_{ilk} x_{i}(t) x_{l}(t)x_{k}(t)+\ldots,
\end{eqnarray}
where $a_{0}$, $a_{i}$, $a_{(i,l)}$,  $\ldots$ are \textit{weight coefficients} whose values are initially unknown.
These coefficients determine the contribution of the corresponding input factors $x_{i}(t)$, $ i = 1,\; 2,\; \ldots,\; M$ or the product of these input factors in Eq.~(\ref{eq_general_polinom}) to obtain a correct value of the output factor $y(t)$. In other words, the weight coefficients  can be considered as a quantitative measure of the significance of the input factor or the corresponding product of the input factors. Obviously, the greater the number of the input factors, the more precisely it is possible to reproduce the target values of the output factor $y(t)$. However, as can be seen from expression~(\ref{eq_general_polinom}), the greater the number
of the input factors, the more complex this expression. In general, in order to be able to formulate
expression~(\ref{eq_general_polinom}), it is necessary that value of $M$ be finite.

Since  expression~(\ref{eq_general_polinom}) contains all the possible combinations of the
products of the input factors, then the problem of finding expression~(\ref{eq_general_polinom})
in its explicit form reduces to finding values of the weight coefficients $a_{0}$, $a_{i}$, $a_{il}$, $\ldots$.

\subsection{Search for the type of regression model as an optimization problem. }

Expression~(\ref{eq_general_polinom}) can be rewritten in a form which will be common for the products of the input factors $x_{1}(t)$, $x_{2}(t)$, $x_{3}(t)$, $\ldots$, $x_{M}(t)$:
\begin{eqnarray} \label{eq_1.2.1}
    y(t)&=&a_{1}\cdot x_{1}^{\delta_{1}^{(1)}(t)} \cdot x_{2}^{\delta_{2}^{(1)}(t)}
    \cdot x_{3}^{\delta_{3}^{(1)}(t)} \cdot \ldots\cdot x_{M}^{\delta_{M}^{(1)}(t)}+ \nonumber \\
    & & + a_{2}\cdot x_{1}^{\delta_{1}^{(2)}(t)}\cdot x_{2}^{\delta_{2}^{(2)}(t)}
    \cdot x_{3}^{\delta_{3}^{(2)}(t)}\cdot \ldots\cdot x_{M}^{\delta_{M}^{(2)}(t)}+ \nonumber \\
    & & + a_{3}\cdot x_{1}^{\delta_{1}^{(3)}(t)} \cdot x_{2}^{\delta_{2}^{(3)}(t)}
    \cdot x_{3}^{\delta_{3}^{(3)}(t)}\cdot\ldots\cdot x_{M}^{\delta_{M}^{(3)}(t)}+\cdots = \nonumber\\
    &=&\sum_{i=1}^{2^{M}}a_{i}\cdot x_{1}^{\delta_{1}^{(i)}(t)}\cdot x_{2}^{\delta_{2}^{(i)}(t)}
    \cdot x_{3}^{\delta_{3}^{(i)}(t)}\cdot\ldots\cdot x_{M}^{\delta_{M}^{(i)}(t)}=    \nonumber \\
    &=&\sum_{i=1}^{2^{M}}a_{i}\prod_{j=1}^{M}x_{j}^{\delta_{j}^{(i)}(t)},
\end{eqnarray}
where $i$ is the order number of the product of the input factors, and $j$ is the index of the input factor in the corresponding product, and the quantity $\delta_ {j}^{(i)}$ characterizes the exponent and can take the values $0$ or $1$, i.e. $\delta_{j}^{(i)} = 0$ or $\delta_{j}^{(i)} = 1$. We introduce the next notation:

\begin{eqnarray} \label{eq_1.2.2}
    \bar{X}_{i}(t) = \prod_{j=1}^{M}x_{j}^{\delta_{j}^{(i)}(t)}= \nonumber \\
    = x_{1}^{\delta_{1}^{(i)}(t)}\cdot x_{2}^{\delta_{2}^{(i)}(t)}\cdot x_{3}^{\delta_{3}^{(i)}(t)}\cdot\ldots\cdot  x_{M}^{\delta_{M}^{(i)}(t)},
\end{eqnarray}

where the factors in this product are written in ascending order of their indices. Then, expression~(\ref{eq_general_polinom}) takes the form
\begin{equation} \label{eq_1.2.3}
    y(t)=\sum_{i=1}^{2^{M}} a_{i} \bar{X}_{i}(t).
\end{equation}
From a mathematical point of view, expression~(\ref{eq_1.2.3}) is the scalar product of two vectors
\begin{equation}
    \label{eq: vector_weights}
    \bar{\bar{a}}=\{a_{1},\; a_{2},\; a_{3}, \ldots, a_{2^M}\}
\end{equation}
and
\begin{equation}
    \label{eq: vector_products}
    \bar{\bar{X}}(t)= \{\bar{X}_{1}(t),\;\bar{X}_{2}(t),\;\bar{X}_{3}(t),\; \ldots,\; \bar{X}_{2^M}(t)\},
\end{equation}
and can be written as
\begin{equation} \label{eq_1.2.4}
    y(t)=(\bar{\bar{a}},\bar{\bar{X}}(t)).
\end{equation}

Then, the search for expression~(\ref{eq_general_polinom}) [or equivalent expression~(\ref{eq_1.2.1})], which would correctly reproduce values of the output factor~$y(t)$, is reduced to the search for the exponents $\delta_{j}^{(i)}$ in the products $\bar{X}_i(t)$. We note that this is a typical optimization problem~\cite{tempbib8, tempbib22,tempbib25,tempbib32}. In the case when the number of the input factors is not large, such problems are solved by the so called \textit{brute force method}~\cite{tempbib9}. However, when there are a sufficiently large number of the input factors, solution to this problem can be obtained using the machine learning methods, for example, using the genetic algorithms~\cite{tempbib4,tempbib23,tempbib26,tempbib28,tempbib29,tempbib30,tempbib31}.

\subsection{Genetic Algorithms. Basic definitions and methodology.}

Genetic algorithms is aimed to solve the optimization problem and are based on the ideas typical for a natural evolutionary process. This allows one to avoid to perform a consequent consideration of all the possible forms of the equation for the output factor $y(t)$ with various combinations of the input factors to find the optimal solution~\cite{tempbib4}. The common GA methodology includes the use of basic notions such as \textit{individual}, \textit{genome} and \textit{population} as well as the two specific operations~\cite{tempbib4}. As applied to solving the problem of finding the optimal regression model for $y(t)$,  we introduce the following definitions:

(i) We define a model polynomial of the form (\ref{eq_1.2.1}) for the output factor $y(t)$ as an \textit{individual}. In general case, the concrete form of an individual $y(t)$ can be specified arbitrarily or in accordance with some rule. It is reasonable to take  a model polynomial for $y(t)$ as an individual, because the this quantity is of the main interest in the given problem and this quantity should be found directly by means of the genetic algorithms~\cite{tempbib10}.

(ii) According to the GA methodology, uniqueness of an individual should be determined by a \textit{genome}. As seen from expression~(\ref{eq_1.2.1}) and Eq.~(\ref{eq_1.2.2}), all input factors are contained in each product~$\bar{X}_{i}(t)$, and these products are included in the model polynomial for the output factor $y(t)$. Account for the concrete input factor (say, the factor $x_j$) in a product $\bar{X}_{i}(t)$ is due to a value in exponent $\delta$ at this factor; recall that this value can be $0$ or $1$. So, for example, when we need to take into account the input factor $x_{j}$ in the product $\bar{X}_{i}(t)$, we should take the exponent $\delta_{j}^{(i)}=1$. If we need to exclude the input factor $x_j$ from consideration, then we take the exponent $\delta$ at the corresponding factor as  $\delta_{j}^{(i)}=0$. Thus, each sequence of products of input factors will be characterized by a unique set of ones and zeros. Then, it is reasonable to define this unique set of zeros and units as a \textit{genome}. To clarify this, we provide the following example. Let there be four input factors $x_1$, $x_2$, $x_3$ and $x_4$. Then, a some product, say,  $\bar{X}_5$ takes the following form:
\begin{equation} \label{eq_1.3.1}
    \bar{X}_5=x_{1}^{0}x_{2}^{1}x_{3}^{0}x_{4}^{1}\equiv x_{2}^{1}x_{4}^{1}\equiv x_{2}x_{4},
\end{equation}
where the set of exponent values $\{0,\;1,\;0,\;1\}$ is a binary notation for the digit $5$, i.e. for the index $i=5$. In other words, a natural correspondence appears between the product index (in the given example with $\bar{X}_5$, this is $i=5$) and the input factors, which are included in this product. Recall that the input factors in expression (\ref{eq_1.3.1}) should be written in order of increasing index.

(iii) By \textit{population} we mean a set of individuals:
\begin{eqnarray} \label{eq_1.3.2}
    y^{(1)}(t)&=&\sum_{i}a_{i}^{(1)}\bar{X}_{i}(t), \nonumber\\
    y^{(2)}(t)&=&\sum_{i}a_{i}^{(2)}\bar{X}_{i}(t), \nonumber \\
    y^{(3)}(t)&=&\sum_{i}a_{i}^{(3)}\bar{X}_{i}(t),\\
        & &\cdots, \nonumber \\
    y^{(s)}(t)&=&\sum_{i}a_{i}^{(s)}\bar{X}_{i}(t), \nonumber \\
        & & \cdots, \nonumber
\end{eqnarray}
where $s$ is the individual index in the population. Each model polynomial $y^{(s)}(t)$ of set~(\ref{eq_1.3.2}) can be considered as some possible solution for the output factor $y(t)$.

In addition, in accordance with the GA methodology, two basic evolutionary operations must be defined: crossover and mutation operations~\cite{tempbib12,tempbib11,tempbib34,tempbib27}.

\begin{figure*}[!tbh] %
\centering
	 \includegraphics [clip,width=180mm]{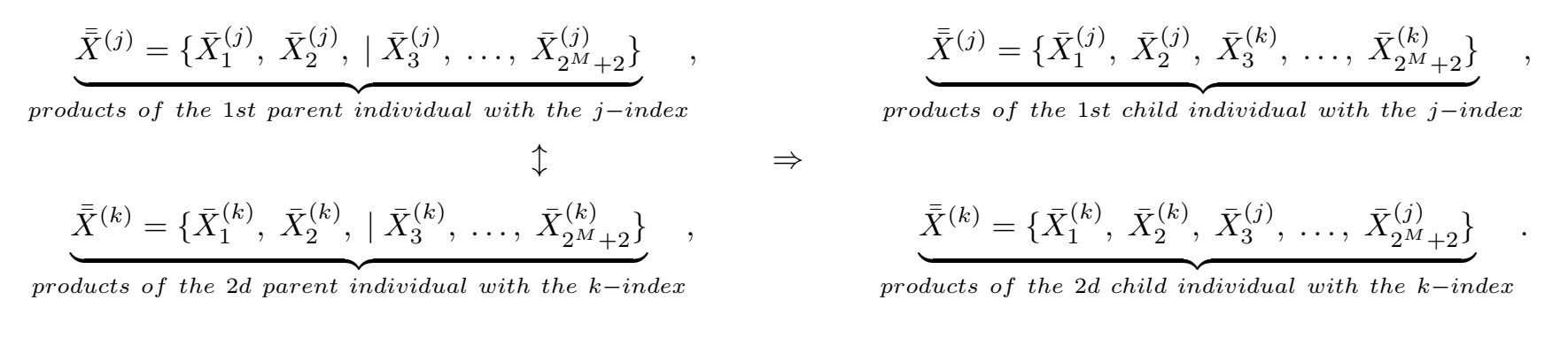}%
\caption{The crossover operation}
\label{fig: 1_1}
\end{figure*}

(iv) We define the \textit{crossover operation} as the operation of obtaining a new pair of the ''child'' individuals by exchanging the right fragments of the vectors $\bar{\bar{X}}^{(j)}$ and $\bar{\bar{X}}^{(k)}$, which belong to two ''parent'' individuals $y^{(j)}$ and $y^{(k)}$ of the same population [see Eqs.~(\ref{eq: vector_products}) and (\ref{eq_1.2.4})]. Here, $j$ and $k$ are the indices of the individuals in the population. Namely, to generate a new pair of individuals, it is necessary to specify the so-called \textit{separation point}, which will be a natural number $n$, and $n\in[1,\;2^M+2]$. Here, the largest value of $n$ is $2^M+2$, because there are $2^M$ various forms of the products $\bar{X}$ of the input factors [see Eq.~(\ref{eq_1.2.1})] and two contributions $\bar{X}_{*}=0$ and $\bar{X}_{0}=1$ in the same population. The point corresponding to this natural number $n$ will divide the vectors $\bar{\bar{X}}^{(j)}$ and $\bar{\bar{X}}^{(k)}$ into left and right parts. By swapping the right-hand sides of these vectors, we get a new pair of the vectors and, thereby, we get a pair of the ''child'' individuals. This operation is illustrated by the following scheme
(see Fig.~\ref{fig: 1_1})

(v) Mutation is an operation due to which a new individual with a changed gene appears~\cite{tempbib13}.
Recall that, in accordance with our methodology, the gene is represented as a set of ones and zeros [see, for example,~(\ref{eq_1.3.1})], and this set characterizes the presence or absence of a factor $x_j$ in a product $\bar{X}_{i}$. Therefore, to obtain a new individual by means of the \textit{mutation}, we need to realize inversion of a random digit in this set: namely, zero inverts to one, or vice versa. So, for example, in the case of the product $\bar{X}_{5}$, that was given above, a possible result of the mutation can be
\begin{eqnarray} \label{eq_1.3.3}
    \bar{X}_5=x_{1}^{0}x_{2}^{1}x_{3}^{0}x_{4}^{1} \; \Rightarrow \; \bar{X}_7=x_{1}^{0}x_{2}^{1}x_{3}^{1}x_{4}^{1}\equiv \nonumber\\ x_{2}^{1}x_{3}^{1}x_{4}^{1}\equiv x_{2}x_{3}x_{4}.
\end{eqnarray}
Here, the product $\bar{X}_7$ appears directly as a result of the mutation with inverse procedure of the exponent for the input factor $x_{3}$.

\subsection{Implementation of GA Methodology. \label{sec: implem_GA}}

The main purpose of the above methodology is searching the most optimal polynomial $y(t)$, which is expressed as~(\ref{eq_general_polinom}) and is able to reproduce the experimental (target) values of the output factor.
Let we have a dataset for the time period $[0,\;T]$. To implement the GA, it is necessary to define two ranges that correspond to \textit{learning sampling} and \textit{test sampling}~\cite{tempbib12,tempbib11}. A rigorous approach requires solving the problem of finding the optimal size of these samplings. It should be noted that if the sampling sizes turn out to be small, then the available data in these samplings will not be enough for the proper statistical analysis, and, therefore, the resulted model for $y(t)$ will not be of a sufficient accuracy.
On the other hand, the larger the sizes of these samplings, the more computing resources are required.
Thus, the solution to this problem implies finding an optimal size of the samplings.
In the simplest implementation, one can set the learning and test samplings of a same size. Moreover, to have a possibility to do a forecast with the data, the overall  time range $[0,\;T]$ can be divided into three ranges of the same size: $[0,\; T/3]$, $ [T/3, \; 2T/3]$ and $[2T/3, \; T]$, which will be associated with learning sampling, test  sampling and \textit{forecast sampling}, respectively.

The first stage of the methodology implies the formation of a trial population of individuals $y^{(s)}(t)$, where $s = 1,2,\ldots$.
Data corresponding to the learning sampling (for example, the time range $[0,\;2T/3]$) will be used to generate a set of model polynomials $y^{(s)}(t)$.
This is a \textit{learning regime}. ''Experimental'' data for the output factor $y(t)$ from the test sampling (for example, from the time range $[2T/3,\;T]$) should be used to test how the resulting models $y^{(s)}(t)$  are able to reproduce the ``experimental'' data. To estimate quantitatively the accuracy of a resulting model, a \textit{matching criterion} $\Delta$  -- e.g., the Fisher criterion, the standard deviation, the coefficient of variation or the coefficient of determination -- should be used. The matching criterion $\Delta$ is computed for each polynomial $y^{(s)}(t)$ of a population, and, thus, the set of $\Delta^{(s)}$ is formed~\cite{tempbib11}.

The second stage of the GA implies numerical calculation of the set of vectors $\bar{\bar{a}}$ [see expression~(\ref{eq_1.2.4})], as well as finding the explicit form of the polynomials $y^{(s)}(t)$ of a population and finding the values of the corresponding criteria $\Delta^{(s)}$.
Based on the found values of the criterion $\Delta$, all the individuals of the population are ranked in order of increasing values of $\Delta$. Namely, the individuals $y^{(s)}$, whose the matching criterion $\Delta$ takes the highest values, are at the end of the set.
Then, the lower half of the ranked population is discarded. Thus, the ranked set of values of $\Delta^{(s)}$ makes a sense of the so-called \textit{fitness function}, which is used in the GA to determine an acceptable solution domain~\cite{tempbib4}.

At the third stage of the algorithm, a population is replenished by means of the crossover operation for the individuals and the mutation applied for the genes, as was described above. We note that the crossover operation applied for a random pair of individuals from the remaining half of the population is carried out until the original size of the population will be restored. The mutation for different pairs of individuals is performed with a certain given probability.

The second and third stages are cyclically repeated until the first polynomial $y^{(s)}$ of the population ranked by the values of the matching criterion will corresponds with the required accuracy to  ``experimental'' data for the output factor $y(t)$.

\begin{table*}[h]
	\caption{Input and output factors characterizing a current lesson}
	\label{tab:1}
	\begin{tabular}{clc}
		\hline
		\hline
Factor \\ notation & Factor Description                                          & Group of factors  \\
		\hline
        \hline
		$x_1$ 	& Factor indicating whether new knowledge was given by teacher in a lesson &    \\
		$x_2$ 	& Number of schoolchildren in a lesson                                   &  \\
		$x_3$	& Factor indicating whether test work was in a lesson                      & Organization \\
		$x_4$ 	& Factor indicating whether independent work was in a lesson               & of a lesson \\
		$x_5$  	& Factor indicating whether demonstration experiment was given             &   \\
		$x_6$ 	& Factor indicating whether laboratory work  was in a lesson               &    \\
        \hline
		$x_7$ 	& Number of schoolchildren who received the assessment ''satisfactory''    & Assessments  \\
        $x_8$   & Number of schoolchildren who received the assessment ''unsatisfactory''  & of academic performance\\
        \hline
        $x_9$   & Have tasks been set for homework?                                         & \\
        $x_{10}$  & Have repetition of theoretical knowledge been given as homework?         &  Homework \\
        $x_{11}$  & Have tasks associated with the search of an additional information       &  \\
                  & been given as homework?                                       &  \\
        \hline
        $x_{12}$  & Noise factor                                                             &  Noise factors \\
        $x_{13}$  & Noise factor                                                             &  \\
        \hline
        $y_1$     & Number of schoolchildren who received                                    &  Output factor \\
                  &     the assessments ''good'' and ''excellent'' in a lesson     &  \\
 		\hline
 		\hline
	\end{tabular}
\end{table*}

\section{Analysis of data characterizing the educational process in the discipline of ''physics'' \label{sec: physics}}

Applying the machine learning methods in an educational process represents a completely new research field, which can be aimed at improving the efficiency and quality of education~\cite{tempbib35}. The interest in using the machine learning algorithms for solving various educational problems is growing, as evidenced by the regular appearance of new studies offering various original approaches~\cite{tempbib14,tempbib15,tempbib17,tempbib16,tempbib18}. In particular, in the review paper~\cite{tempbib14} it is proposed to use these methods to predict student performance.  At the same time, it is proposed in Ref.~\cite{tempbib14} to use these methods to make various forecasts on the basis of an information relating to the social, age and psychological features of students. It is noteworthy that the ideas given in Ref.~\cite{tempbib14} are consistent with the ideas proposed before in Refs.~\cite{tempbib15, tempbib17}.
It is important to note that  applying the machine learning methods in a completely different context for the educational process was discussed in Refs.~\cite{tempbib16, tempbib18}. Namely, as demonstrated in Refs.~\cite{tempbib16, tempbib18}, it is possible to use the machine learning methods on the basis of the so-called flexible tree-based algorithms to provide an accurate and maximally objective assessment of the schoolchildren and students performance on the basis of the test results. The corresponding results were obtained from a comparative analysis of schoolchildren/student performance from nine different countries.

In this work, we are implementing the GA methodology presented in Section~\ref{sec: method} to analyze the data for the educational process related with the physics discipline in the concrete secondary school in Kazan (Russia).
The analyzed temporal period is covered the data for one and a half academic years. This period includes $97$ lessons for the humanitarian designated class (H-class) and the $146$ lessons for the class  designated to natural sciences (NS-class). The data contained the following information, namely, (i)  the information about the students assessments for each lesson, (ii) the information about whether the test work, independent work,  laboratory work was in a lesson, whether a demonstration experiment was given during a lesson, and also (iii) the information about the type of hometask for the schoolchildren.

\begin{figure}

	\centering 
	\includegraphics [clip,width=80mm]{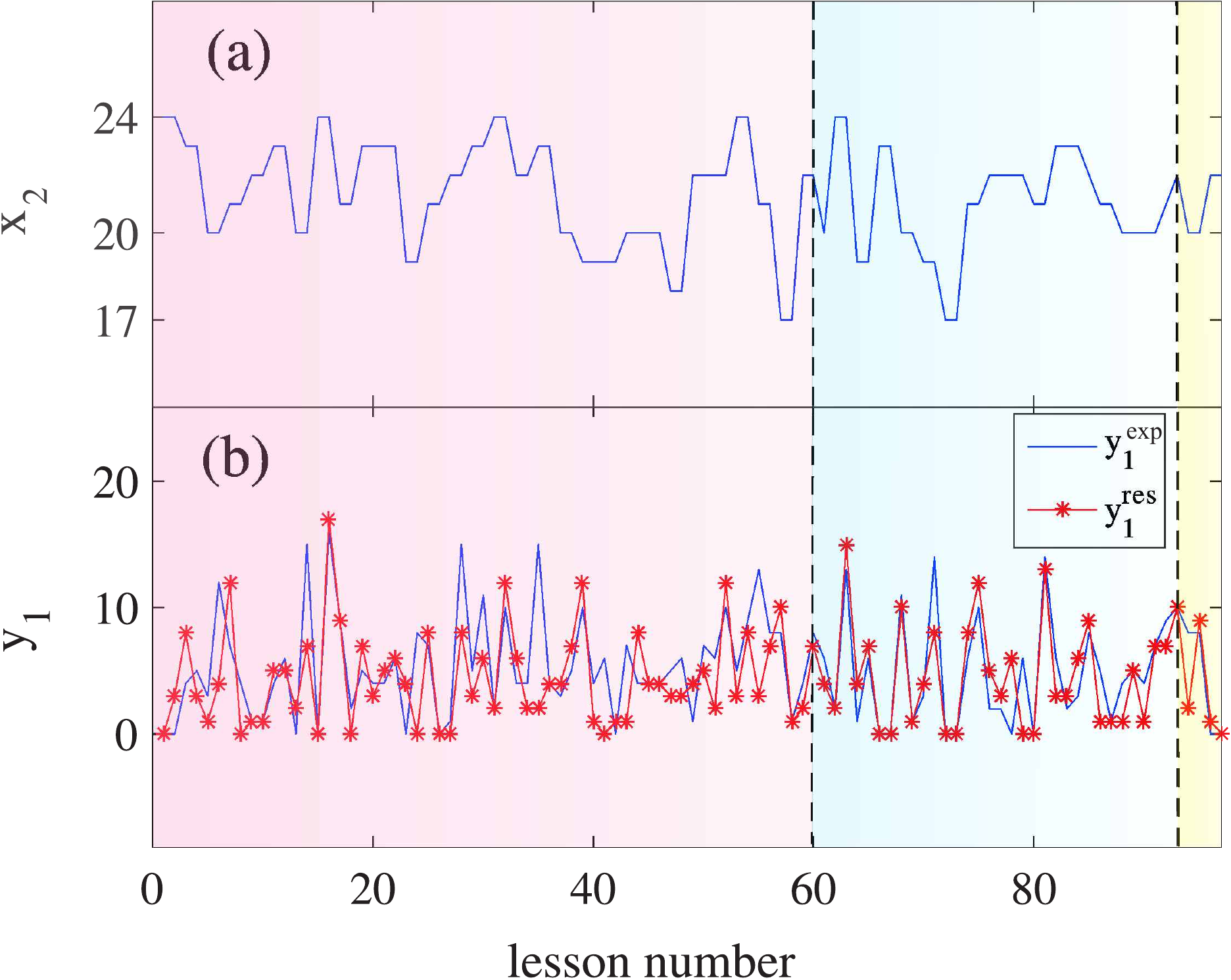} 
\caption{Time-dependent factors for the H-class:  input factor $x_2$ indicating the number of students in a lesson, and output factor $y_1$ which indicate the number of schoolchildren who received the assessments ``good'' and ``excellent'' in a lesson. Note that the number of schoolchildren in the full class is $24$. Solid (blue) lines represent experimental data; red stars connected by solid (red) line is a model result with Eq.~(\ref{eq_model}). The learning, test and forecast samplings are separated by vertical dashed lines. } \label{fig: 1}
\end{figure}
The main aim of this study is  to evaluate the efficiency of teaching physics in these concrete lessons at this school.  As a result of this, the factor $y_1$ characterizing the number of schoolchildren received  the assessments ''good'' and ''excellent'' during a lesson was chosen as an output factor.   All other factors were assigned to the input factors (for detail, see Tab.~\ref{tab:1}). As seen from this table, all the input factors $x_i$, $i=1,\;2,\;\ldots,\;11$, can be grouped as the follows: (a) the group characterizing the organization of a lesson, (b) the group characterizing the assessments of academic performance and (c) the group that characterizes the type of homework. Such a division of the input factors into the groups may make it possible to understand how the certain groups of factors influence the academic performance. In addition, it is worth noting that we artificially introduced two additional noise factors, $x_{12}$ and $x_{13}$, which represented the arrays of random natural numbers. This was done in order to verify the accuracy of the algorithm used. So, if the algorithm works correctly, then the algorithm should recognize that these noise factors are not associated with the physics teaching data. Note that the factors $x_1$, $x_3$, $x_4$, $x_5$, $x_6$, $x_9$, $x_{10}$ and $x_{11}$ can only take two wordings: ''yes''  or "no". In our numerical analysis, we apply the next correspondence between these wordings and the numeric denotations: $0$ -- ''no'' and $1$ -- ''yes''.

\begin{figure*}
\centering
	 \includegraphics [width=150mm] {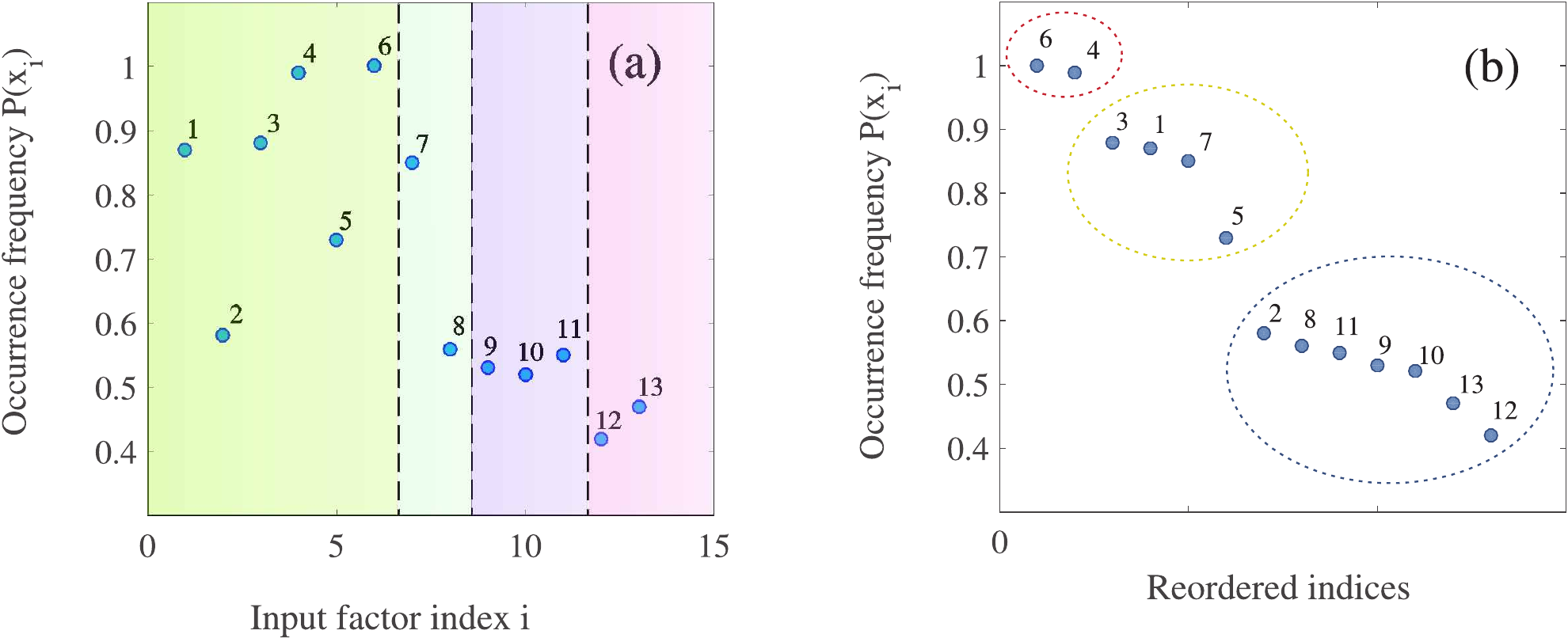}
\caption{(a) Occurrence frequency $P(x_i)$ of the considered input factor $x_i$ depending on the index of this factor. Note that these frequencies were obtained from the procedure of the regression model formation [see Eq.~(\ref{eq_model})]. Various groups of the input factors are separated by vertical dashed lines. (b) Occurrence frequencies presented in order of decreasing their values.
}
\label{fig: 2}
\end{figure*}
Let us consider the results of analysis for the H-class.
According to the algorithm presented in Section~\ref{sec: implem_GA}, on the basis of the data corresponding to the H-class the following regression model was obtained:
\begin{eqnarray} \label{eq_model}
  y_1(t) &=& a^{(1)} + a^{(2)}x_5(t)x_8(t) + a^{(3)}x_3(t)x_{11}(t) + \nonumber \\
         & & + a^{(4)}x_3(t)x_5(t) + a^{(5)}x_8(t)x_9(t) +  \nonumber \\
         & & + a^{(6)}x_{11}(t) + a^{(7)}x_1(t)x_3(t) + a^{(8)}x_8(t) +\nonumber \\
         & & + a^{(9)}x_9(t)+ a^{(10)}x_4(t)x_6(t)+ \nonumber \\
         & & + a^{(11)}x_3(t)x_6(t) + a^{(12)}x_1(t)x_8(t)+ \nonumber\\
         & & + a^{(13)}x_9(t)x_{11}(t)+ a^{(14)}x_{11}(t)+ \nonumber \\
         & & + a^{(15)}x_6(t)x_7(t) + a^{(16)}x_{10}(t)x_{11}(t)+ \nonumber\\
         & & + a^{(17)}x_1(t)+ a^{(18)}x_6(t)x_{11}(t)+ \\
         & & + a^{(19)}x_1(t)x_2(t) + a^{(20)}x_6(t)x_9(t) + \nonumber\\
         & & + a^{(21)}x_9(t)x_{10}(t) + a^{(22)}x_2(t)x_3(t) + \nonumber\\
         & & + a^{(23)}x_5(t)x_6(t) + a^{(24)}x_1(t)x_7(t) + \nonumber\\
         & & + a^{(25)}x_2(t)x_{13}(t) + a^{(26)}x_4(t)x_{11}(t)+ \nonumber\\
         & & + a^{(27)}x_{13}(t) + a^{(28)}x_5(t)x_{11}(t) + \nonumber \\
         & & + a^{(29)}x_4(t)x_9(t) + a^{(30)}x_3(t) + \nonumber \\
         & & + a^{(31)}x_3(t)x_8(t) + a^{(32)}x_9(t)x_{12}(t) + \nonumber \\
         & & + a^{(33)}x_7(t) + a^{(34)}x_{10}(t)x_{13}(t)+\nonumber \\
         & & + a^{(35)}x_2(t)x_{10}(t) + a^{(36)}x_7(t)x_{13}(t) + \nonumber \\
         & & + a^{(37)}x_2(t)x_9(t) + a^{(38)}x_5(t)x_{13}(t)+ \nonumber \\
         & & + a^{(39)}x_7(t)x_{11}(t) + a^{(40)}x_1(t)x_5(t) + \nonumber \\
         & & + a^{(41)}x_{11}(t)x_{12}(t) + a^{(42)}x_2(t)x_5(t). \nonumber
\end{eqnarray}

This regression model was obtained as follows. Since the total time domain was $T = 97$ corresponding to $97$ lessons, then we have defined the learning sampling of the size $[0, \; 60]$ lessons and the test sampling of the size  $[60, \; 92]$  lessons. Further, the sampling of the size $[92, \; 97]$  lessons was taken for the forecast (see Fig.~\ref{fig: 1}). It is remarkable that, as seen from Fig.~\ref{fig: 1}, the obtained regression model~(\ref{eq_model}) reproduces correctly the experimental data of the output factor $y_1(t)$ for the test sampling as well as for the forecast sampling.

As a result of constructing the regression model for $y_1(t)$ [see Eq.~(\ref{eq_model})], the occurrence  frequency $P(x_i)$ of each the input factor $x_i$, $i=1,\;2,\;\ldots,\;13$, was obtained. In fact, the occurrence frequency, normalized to unity, indicates the probability of the appearance of a certain input factor during the procedure for finding the model polynomial $y_1(t)$. The found frequencies for the input factors are presented in Fig.~\ref{fig: 2}(a), and the number near each symbol (circle) indicates the index of the input factor corresponding the occurrence frequency. Obviously, the higher the occurrence frequency of the input factor $x_i$, the more significant this factor for the behaviour of the output factor $y_1(t)$. Fig.~\ref{fig: 2}(b) shows the same occurrence frequencies in order of decreasing their values. As seen, the frequencies are grouped into three separate regions. The region with the highest frequency values located in the upper left corner in Fig.~\ref{fig: 2}(b) contains the input factors $x_6$ and $x_4$ that have the \textit{greatest impact} on the behaviour of the output factor $y_1(t)$. Recall that these factors characterize whether the laboratory and independent works were given in a lesson.
The second group of the input factors, which can be designated as \textit{less significant} factors, is located in the middle of the graph shown in Fig.~\ref{fig: 2}(b). This group includes the input factors that provide the following information: whether the test work was given in a lesson ($x_3$), whether new knowledge was given by teacher in a lesson ($x_1$), the number of schoolchildren who received the assessment ''satisfactory'' in a lesson ($x_7$), and, finally, whether the demonstration experiment was given in a lesson ($x_5$). Other factors located in the lower right corner in Fig.~\ref{fig: 2}(b) appears to be the \textit{least significant}. It is important to note that the similar results were obtained from the analysis for the NS-class.

Based on the performed analysis, we come to the following interesting conclusions:

(i) It turned out that the factors characterizing the \textit{organization of the lesson}, namely, the factors associated with test work, independent work and laboratory work, most strongly affect the academic performance of schoolchildren in physics in this secondary school.

(ii) It was obtained the nontrivial result that the academic performance is not strongly dependent on \textit{how many schoolchildren} attend in a lesson. Thus, there is a contradiction with the well known opinion that the less students are present at the lesson, the higher their academic performance should be. Nevertheless, it is important to take into account that the non-trivial result obtained by us in this study follows from the data for the concrete discipline in the concrete school, and, therefore, this conclusion is relevant only to this data.

(iii) As turned out, the GA algorithm allows one to find the mathematical model for the academic performance, on the basis of which a realistic forecast for the nearest future can be done.

(iv) Unexpectedly, the group of factors related with the \textit{homework}  has a little effect on the student academic performance.

(v) Finally, the \textit{noise factors} $x_{12}(t)$ and $x_{13}(t)$ have no impact on the academic performance. This conclusion follows directly from the obtained distribution $P(x_i)$ presented in Fig.~\ref{fig: 2}(b):  these noise factors have the lowest values of the occurrence frequency and are at the right bottom part of the distribution. Consequently, the GA algorithm recognizes these factors as not related to the real input factors for the considered complex system. This is evidence that the algorithm is resistant to artificially introduced ''disturbances''.

\section{Conclusion}

In this study, we present the approach for analyzing the evolution of a complex system. The approach is based on the GA-algorithms and allows one to solve the specific problems associated with  identifying the significant factors by means of which it is possible to drive by a complex system, with optimizing the dynamics of a complex system, and with  predicting the evolution of a system.

As an example, we have considered the data associated with teaching the discipline ``Physics'' in the concrete secondary school in Kazan (Russia). On the basis of the regression model generated for the schoolchildren academic performance, the some results were obtained. In particular, as shown, it is possible to identify the main factors affecting the educational process, namely, teaching the discipline ``Physics'' in a concrete school. In addition, as it turned out, it is possible to make a probabilistic forecast of student performance for some immediate period of time, provided that the general educational scheme for the discipline remains unchanged. The results generated by means of the GA-algorithm for the education data can be applied to yield the recommendations to improve the efficiency of a specific educational process. It should be noted that the original approach presented in this study can be applied to any other school discipline. Finally, this approach can be generalized to solve the problems associated with improving the efficiency of the educational institutions as a whole.

\section*{Acknowledgements}
\vspace{\baselineskip} This work is supported by the Russian Foundation for Basic Research (the project No.~$18$-$02$-$00407$).

%

\end{document}